\documentclass{article}

\usepackage{arxiv}

\usepackage[utf8]{inputenc} 
\usepackage[T1]{fontenc}    
\usepackage{hyperref}       
\usepackage{url}            
\usepackage{booktabs}       
\usepackage{amsfonts}       
\usepackage{nicefrac}       
\usepackage{microtype}      
\usepackage{lipsum}
\usepackage{tablefootnote}
\usepackage{amsmath}
\usepackage{multirow}

\usepackage{graphicx}
\graphicspath{ {./images/} }

\title{Using Machine Learning and Alternative Data to Predict Movements in Market Risk}

\author{
  Thomas Dierckx\\
  Department of Statistics\\
  Department of Computer Science\\
  KU Leuven, Belgium\\
  \texttt{thomas.dierckx@cs.kuleuven.be} \\
   \And
    Jesse Davis  \\
  Department of Computer Science\\
  KU Leuven, Belgium\\
  \texttt{jesse.davis@cs.kuleuven.be} \\
   \AND
   Wim Schoutens \\
   Department of Statistics \\
   KU Leuven, Belgium \\
   \texttt{wim.schoutens@kuleuven.be} \\
}

\begin{document}
\maketitle

\begin{abstract}
Using machine learning and alternative data for the prediction of financial markets has been a popular topic in recent years. Many financial variables such as stock price, historical volatility and trade volume have already been through extensive investigation. Remarkably, we found no existing research on the prediction of an asset's market implied volatility within this context. This forward-looking measure gauges the sentiment on the future volatility of an asset, and is deemed one of the most important parameters in the world of derivatives. The ability to predict this statistic may therefore provide a competitive edge to practitioners of market making and asset management alike. Consequently, in this paper we investigate Google News statistics and Wikipedia site traffic as alternative data sources to quantitative market data and consider Logistic Regression, Support Vector Machines and AdaBoost as machine learning models. We show that movements in market implied volatility can indeed be predicted through the help of machine learning techniques. Although the employed alternative data appears to not enhance predictive accuracy, we reveal preliminary evidence of non-linear relationships between features obtained from Wikipedia page traffic and movements in market implied volatility. 
\end{abstract}

\keywords{Implied Volatility \and Machine Learning \and Alternative Data}

\section{Introduction}

To predict the stock market, statisticians generally relied on econometric methods. However, recent literature \cite{PstSPUEMaODS} shows that machine learning models typically outperform statistical and econometric models (\cite{SmodAMPUDDS, Hsu2016, Meesad2013, Patel2015b, Zhang2009}). Machine learning models typically provide more flexibility by requiring less distributional assumptions about the data \cite{Zhang2009}. Moreover, they are able to recognize patterns in time series data more easily \cite{Meesad2013} and can even be combined to reduce over-fitting and further improve performance \cite{Patel2015b}. In addition, with the recent paradigm shift in machine learning towards deep learning, researchers can extract features and model non-linear correlations without relying on econometric assumptions and human expertise \cite{CDDLAfMFMP}. Selecting an appropriate model alone is never sufficient since predictive performance will significantly be determined by the employed data. Most studies have traditionally focused on quantitative market data (e.g. \cite{Patra2009, Saad1998, Roman1996, Jia2016, Chiang2016, Chong2017, Goumatianos2017}). Interestingly, since the inception of Web 2.0, research has been exploring numerous alternative data sources to improve predictive accuracy. These novel sources, such as social media and online news, exert a positive impact on information diffusion and thus affect the market and its investors in various ways. Although it is not clear whether we are rather emotional than rational beings, it is not inconceivable that investors can act emotionally and are influenced by these alternative data sources. This is in line with proponents of behavioural science, where investment decisions are influenced by investor sentiment (\cite{Baker2006, Clark-Murphy2005, Tauni2017}). Moreover, studies have already shown that investors often exhibit herd behaviour (\cite{Scharfstein1990, Chang2000, Ivkovic2007}). Consequently, fueled by advancements in Natural Language Processing (NLP), the last few years have seen an explosion in research on quantifying the synergy between online media and stock markets. On top of that, due to the surge in popularity of certain data analysis techniques and increasing computational power, it has never been easier to examine vast amounts of features for more accurate predictions.

A wealth of financial market variables have already been subject to prediction attempts by research using both machine learning and alternative data. Most research focuses on stock price prediction, seen as the \textit{holy grail} within finance (e.g. \cite{MTPuSALLaPF, MTPuSALLaPF, UTTNfSMA, CtWoCaTAfFMPUDRSE, LtCWADLFfNoSTP, SmodAMPUDDS}). However, other popular market variables such as historical volatility (\cite{Atkins2018, Oliveira2017, Rekabsaz2017, YoungKim2018} and trade volume \cite{Oliveira2017} have also been considered. Remarkably, we found no research that explored the use of machine learning, not to mention alternative data sources, to predict the market implied volatility of assets. Derived from option prices, this variable is deemed to be one of the more important parameters in the world of derivatives. In contrast to historical volatility, this forward-looking measure indicates how much risk the market \textit{expects} a certain asset to exhibit in the coming period. As this variable is paramount for the pricing of options, the ability to predict its movements would be advantageous for the practice of asset management and market making alike. Consequently, the research focus in this paper is twofold. First we investigate whether we can use machine learning algorithms to predict if an asset's market implied volatility will have moved up or down by the end of the next trading day. Second, we examine whether the introduction of alternative data sources has any positive impact on predictive accuracy.

\section{Preliminaries}
Using historical quantitative market data for stock prediction has been around for decades. Interestingly, since the inception of Web 2.0, the search for predictive information from alternative data sources has expanded notably. A tremendous amount of research has been published on quantifying the interplay between online media and stock markets. As a short literature study, Section \ref{interplay_section} rehashes the most important recent findings based on recent surveys (\cite{Survey1, Survey2, Survey3, Survey4}). Subsequently, Section \ref{implied_section} elaborates on the concept of market implied volatility.

\subsection{The interplay between the stock market and alternative data} \label{interplay_section}

A variety of alternative sources have been considered to aid in stock market prediction thus far \cite{Survey3, Survey4}. Most studies have focused on Twitter because of its popularity and easy accessibility (e.g. \cite{Survey1_11, Si2013}), but also discussion boards (e.g. \cite{Nguyen2015, Survey1_9}) and news sources such as The Wall Street Journal, Financial Times and Thomson Reuters (e.g. \cite{Yoshihara2016, Ding2015}) have been considered. Corporate disclosures and financial reports are published only a few times a year and have received somewhat less attention, but are gaining momentum (e.g. \cite{Chan2011, Groth2011}). Moreover, surveys \cite{Survey1} and \cite{Survey4} note that Google Trends and Wikipedia page views are also being considered as predictive sources in recent years (e.g. \cite{SmodAMPUDDS, PstSPUEMaODS}]). Most of the aforementioned sources are mostly comprised of textual data rich in information. In order for this type of content to be useful for machine learning algorithms, transformation into a machine interpretable form, such as a scalar or tensor, is necessary. However, extracting all relevant information contained in text remains a challenging research topic to this day \cite{Survey1}.

Over a decade worth of research is evidently accompanied by a wealth of findings. Many data sources have been investigated for their predictive power, but one that stands out in particular is Twitter. This platform, due to its popularity, accessibility and compact content is significantly more investigated than any other alternative data source. Table \ref{tab:tweets} presents the main findings on Twitter and shows empirical evidence for its predictive power. However, not all research was in favor for the predictive power of tweets. For example, \cite{Brown2012} only found a weak statistical correlation between Twitter sentiment and the S\&P500's closing price. Overall, \cite{Survey2} found 28 papers supporting the predictive power of Twitter sentiment, and 5 papers with mixed or non-supportive results. This suggests that there is indeed at least some supportive evidence for the predictive power of investor sentiment in the context of capital markets. Moreover, \cite{Survey1_17} discovered that social media had a stronger relationship with stock performance than any other alternative data source, and that the impact of different types of sources varied widely. Table \ref{tab:others} presents the most important findings from studies using other types of alternative data such as discussion boards, news articles, Wikipedia or Google Trends as predictors. Although discussion boards and news articles should also contain an abundance of information, there is significantly less research published. 

Once interesting data sources are identified and processed into (hopefully predictive) features, a predictive model needs to be fitted to the data. \cite{Survey3} notes that Regression and Support Vector Machines have been popular methods in the past decade. However, with the recent paradigm shift in machine learning towards deep learning, researchers are starting to favor deep learning methods. Although these methods are prone to overfitting, they can extract features and model non-linear correlations without relying on econometric assumptions and human expertise \cite{CDDLAfMFMP}. Note that deep learning methods generally work best when there is an abundance of training examples available, a characteristic often not present in the financial domain. Which category of algorithms is ultimately most appropriate is still considered an open question \cite{Survey3}. 

\begin{table}[!ht]
\caption{Key Twitter findings throughout the years retrieved from surveys \cite{Survey1} and \cite{Survey2}.}
  \centering
  \begin{tabular}{lll}
    \toprule
    Finding & Year & Reference \\
    \toprule

    The degree to which companies are jointly mentioned is correlated to the co-movement \\
    of these stocks. & 2011 & \cite{Sprenger2011} \\
    \midrule
    Emotions such as "hope", "fear" and "worry" in tweets are negatively related to \\
    stock market indexes and positively correlated to the volatility index .\ & 2011 & \cite{Zhang2011} \\
    \midrule
    Twitter sentiment can be predictive for intraday exchange rates.  & 2013 & \cite{Papaioannou2013} \\
    \midrule
    Sentiment polarity extracted from users with many followers is associated with \\
    abnormal returns on the same day for S\&P 500 stocks.  & 2014 & \cite{Sul2014} \\ 
    \midrule
    Twitter sentiment affects abnormal returns during  peaks of Twitter volume. & 2015 & \cite{Ranco2015} \\
    \midrule
    Existence of a nonlinear causal relationship between Twitter investment on \\
    stock returns on DJIA. & 2015 & \cite{Souza2015}\\ 
    \midrule
    Only negative emotion has significant impact on stock returns. & 2015 & \cite{Risius2015} \\ 
    \midrule
    Similar firms, clustered based on firm-specific microblogging metrics, have higher \\
    co-movements than those in the same industry. & 2015 & \cite{Liu2015} \\ 
    \midrule
    Direct relationship between an IPO its Twitter sentiment and first trading day returns.  & 2016 & \cite{Liew2016} \\ 
    \midrule
    Twitter sentiment can be used to predict the price trend of GOOGL, AAPL and FB. \\
    In addition,  tweet volume has a strong impact on both price and trend.& 2016 & \cite{Corea2016} \\
    \midrule
    Daily bullish percentage extracted from Twitter helps explain excess returns even when \\
    the traditional factors used in asset pricing models are considered. & 2017 & \cite{Liew2017} \\
    \midrule
    Volatility sentiment on social media contains information regarding future stock volatility. & 2017 & \cite{Karagozoglu2017} \\

    \bottomrule
  \end{tabular}
  \label{tab:tweets}
\end{table}

\begin{table}[!ht]
\caption{Key findings from using discussion boards, news articles, Wikipedia and Google Trends based on \cite{Survey1} and \cite{Survey4}.}
  \centering
  \begin{tabular}{lll}
    \toprule
    Finding & Year & Reference \\
    \toprule
    Fraction of negative words in firm-specific news predicts low firm earnings & 2008 & \cite{Survey1_44} \\
    \midrule
    Stocks with no media coverage earned higher returns & 2009 & \cite{Survey1_20}\\
    \midrule
    Connection between sentiment of forum posts and stock indices, volumes and volatility. & 2007 & \cite{Survey1_9} \\
    \midrule
    Page views of articles on Wikipedia relating to companies or financial topics increases \\before stock market declines. & 2013 & \cite{Survey1_50} \\
    \midrule
    Google search volumes for keywords related to financial market increases before \\ stock market declines.& 2013 & \cite{Survey1_57} \\ \midrule
    Movements in financial markets and movements in financial news are intrinsically\\ interlinked & 2013 & \cite{Survey1_74}\\
    \midrule
  Opinions expressed on Seeking Alpha have significant stock returns forecasting ability & 2014 & \cite{Chen2014}\\
    \bottomrule
  \end{tabular}
  \label{tab:others}
\end{table}

\subsection{Market Implied Volatility} \label{implied_section}
In the world of derivatives, options are one of the most prominent types of financial instruments available. As sellers of options are exposed to risk for the duration of the contract, they want to be properly compensated. To measure this risk, the expected price fluctuations of the underlying asset are considered over the course of the option contract. This measure is better known as implied volatility and varies with the strike price and duration of the option contract. Interestingly, when considering the selection of implied volatilities for different strike prices and fixed duration, a more general measure called market implied volatility can be extracted. A famous example of this generalized measure is the CBOE Volatility Index, better known as the VIX. 

VIX is a measure of expected price fluctuations in the S\&P 500 Index options over the next 30 days. It's famously known as the \textit{fear index} and is considered a reflection of investor sentiment on the market. Interestingly, the calculation for the VIX for term $T$ (Equation \ref{eq:vix} - taken from the VIX white paper \cite{VIX}) can be applied to any asset with available options. Although this measure can be calculated for any arbitrary term, the duration of the option contracts will seldom match. To overcome this obstacle, the market implied volatility is calculated for the option contracts expiring right before and after the desired target date. We then linearly interpolate between these two measures to correspond with the desired term, as outlined in \cite{VIX}. Formally, VIX is defined as:

\begin{equation}
  \mathit{VIX} = 100 \times \sqrt{\frac{2}{T} \sum\limits_{i} \frac{\Delta K_{i}}{K_{i}^{2}} e^{RT} Q(K_{i}) - \frac{1}{T} \bigg[ \frac{F}{K_{0}} - 1 \bigg]^{2}}
  \label{eq:vix}
\end{equation}
\noindent where:
\begin{description}
\item[T] is time to expiration
\item[F] is the forward index level derived from the index option prices
\item[K_{0}] is the first strike below the forward index level F
\item[K_{i}] is the first strike price of the $i^{th}$ out-of-the-money option; a call if $K_{i}>K_{0}$ and a put if $K_{i}<K_{0}$; both put and call if $K_{i}=K_{0}$
\item[\Delta K_{i}] is the interval between strike prices
\item[R] is the risk-free rate to expiration
\item[Q(K_{i})] is the midpoint of the bid-ask spread for each option with strike $K_{i}$
\end{description}


\section{Methods}
Recent research focusing on stock price prediction using machine learning and alternative data sources has reported some considerable successes (\cite{PstSPUEMaODS, SmodAMPUDDS, SMPvMSMIL, MTPuSALLaPF, UTTNfSMA, CtWoCaTAfFMPUDRSE, LtCWADLFfNoSTP, CNNP:CNNBSMPUSDS}). Motivated by the novelty of our research, and the promising nature of alternative data, we investigated whether we can predict if an asset's market implied volatility will have moved up or down by the end of the next trading day. In addition to market data, we employed Wikipedia page traffic and Google News statistics as alternative data. Section \ref{data_acquisition} describes in detail how we performed our data acquisition and constructed additional features. Section \ref{ml} then details for which machine learning algorithms we opted, and how we preprocessed our data in order to maximize performance.

\subsection{Data acquisition and feature generation} \label{data_acquisition}
In this paper, we focus on predicting whether AAPL's market implied volatility at the end of the next day will have moved up or down. We considered a 24 month trading period from January 1, 2016 till December 31, 2017 for which three different data sources were utilized. First, we collected publicly available market data from Yahoo Finance including the stock's daily trading volume and daily opening, high, low and closing prices. We obtained the stock's historical market implied volatility by applying the VIX formula (Equation \ref{eq:vix}) on personal end-of-day option data. The second data source is Google News, a source that aggregates news and blog posts relevant to provided keywords. Google News allows for queries on how many publications containing certain keywords were made on any given day. We scraped this information and built a time series of daily news counts for AAPL.
The third and last data source is Wikipedia, which provides daily visitor statistics per page. We retrieved this data via the Wikimedia API, and built a time series of daily page views for the Apple Inc page. In total, eight different features are obtained for each trading day from three different types of data sources (Table \ref{tab:table_data1}).

\begin{table}[!ht]
 \caption{Summary of considered data sources and corresponding original features (daily granularity). }
  \centering
  \begin{tabular}{cccc}
    \toprule
    Market Data  & Option Data & Google News     & Wikipedia \\
    \midrule
    OHCL  & Market Implied Volatility & News Counts  & Page Views    \\
    Volume &   &  &     \\

    \bottomrule
  \end{tabular}
  \label{tab:table_data1}
\end{table}

\begin{table}[!ht]
 \caption{Summary of considered data sources and corresponding original features (daily granularity). }
  \centering
  \begin{tabular}{ccc}
    \toprule
    Market Data   & Google News     & Wikipedia \\
    \midrule
    OHCL  & News Counts  & Page Views    \\
    Volume    &  &     \\
    Market Implied Volatility         &  &     \\

    \bottomrule
  \end{tabular}
  \label{tab:table_data1}
\end{table}

Influenced by the work done in \cite{PstSPUEMaODS} and \cite{SmodAMPUDDS}, we employed a set of technical analysis tools with varying parameters to generate additional predictors on the aforementioned features. Table \ref{tab:table_data2} summarizes our process, where each part of the table is used on a different subset of the eight original features. The techniques listed in the top part (1) are applied only on \textit{market implied volatility}, \textit{news counts} and \textit{page views} yielding 60 new features. The techniques in the middle (2) are applied to the previous features and \textit{close price} yielding eight new features. The last part is solely applied on \textit{close price}, yielding two new features. Note that we replicate the feature generation strategy from \cite{SmodAMPUDDS}, and that daily volume, daily high and daily low are not used for additional feature generation. In total 78 features are obtained for each trading day, yielding 486 feature vectors in total.

\begin{table}[!ht]
 \caption{Summary of selected feature generation techniques with corresponding parameters. The top portion is used on market implied volatility, page views and news count features. The middle is used on the previous features and close price. The last portion is solely used on close price \cite{SmodAMPUDDS}. }
  \centering
  \begin{tabular}{cll}
    \toprule
    & Technique & Parameters \\
    \midrule
     \multirow{10}{*}{(1)} & Moving Average (MA) & $n \in \{3, 5, 10\}$ \\
    & Moving Average Move (MA\_Move) & $n \in \{3, 5, 10\}$ \\
    & Exponential Moving Average (EMA) & $n \in \{3, 5, 10\}$ \\
    & Exponential Moving Average Move (EMA\_Move) & $n \in \{3, 5, 10\}$ \\
    & Rate Of Change (ROC) & $n \in \{5\}$ \\
    & Rate Of Change Move (ROC\_Move) & $n \in \{5\}$ \\
    & Disparity Index & $n \in \{3, 5\}$ \\
    & Disparity Index Move & $n \in \{3, 5\}$ \\
    & Momentum1 \tablefootnote{\label{momentumfnote} See \cite{SmodAMPUDDS} for details on the momentum1 and momentum2 calculations. } & $n \in \{5\}$ \\
    & Momentum2 \tablefootnote{ See footnote \ref{momentumfnote}. } & $n \in \{5\}$ \\
    \midrule
     \multirow{2}{*}{(2)} & Relative Strength Index (RSI) & $n \in \{14\}$ \\
    & Relative Strength Index Move (RSI\_Move) & $n \in \{14\}$ \\
    \midrule
     \multirow{2}{*}{(3)} & Williams \%R  & $n \in \{14\}$ \\
    & Stochastic Oscillator \tablefootnote{ Stochastic Oscillator by George Lane}  & $n \in \{14\}$ \\
    \bottomrule
  \end{tabular}
  \label{tab:table_data2}
\end{table}

To construct the target feature, we define the next day's difference in market implied volatility on day $i$ as $y_{i}^{*} = (\mathit{ivolatility}_{i+1}-\mathit{ivolatility}_{i})$ where $\mathit{ivolatility_{i}}$ denotes the end of day market implied volatility on day $i$. We consider a move to be upwards whenever $y_{i}^{*} > 0$ and downwards whenever $y_{i}^{*} \leq 0$. The final target feature is therefore a binary feature obtained by applying Case Equation \ref{eq:target}.

\begin{equation}
  y_{i}=\begin{cases}
    1, & \text{if $y_{i}^{*}>0$}.\\
    0, & \text{otherwise}.
  \end{cases}
  \label{eq:target}
\end{equation}

\subsection{Machine Learning} \label{ml}
We examined the effectiveness of Logistic Regression, Support Vector Machines (SVM) and Boosting Machines for predicting movements in market implied volatility on the AAPL stock. Although Logistic Regression is a linear model, its speed and interpretability makes it warranted to try \cite{AMToCDiSMPaMIF}. Support Vector Machines have been a popular technique for stock market prediction using alternative data sources (e.g. \cite{SmodAMPUDDS, PstSPUEMaODS, SMPvMSMIL, ESMPwECHMMoMSD}). Trained with the non-linear Radial Basis Function, SVMs are able to fit to both linear and non-linear separable data. Lastly, Boosting Machines such as AdaBoost and XGBoost have surged in popularity last few years and have also been employed in stock market prediction using alternative data sources \cite{PstSPUEMaODS}. Note that when no parameters are mentioned in this or subsequent sections, the standard Python Sklearn library parameter configurations apply.

Feeding raw data to machine learning algorithms is seldom a good idea. One of the main problems when dealing with time series data is the presence of seasonality and trends. We therefore identify non-stationary features with the Augmented Dickey-Fuller test and transform said features into features representing their first difference. Some machine learning algorithms, such as Logistic Regression and Support Vector Machines, also require features to be standardized. We therefore  standardized all features to have zero mean and unit variance before using said models. 

To combat the abundance of features and the associated curse of dimensionality, we used a smaller subset of 29 features for training the Logistic Regression and SVM models. We obtained this smaller subset by fitting a standard AdaBoost model to the data. Tree-based methods like AdaBoost produce a list of relative variable importances, from which we extracted features with a higher than average importance. Aside from using AdaBoost as feature selection tool, we also used the model for prediction. Because forest classifiers such as AdaBoost are robust to large feature spaces and scaling issues, we do not perform standardization or feature selection prior to using this model for classification. 

Note that we consciously refrained from using deep learning techniques due to the lack of data points. Random Forests were also not considered because the algorithm's random sampling clashes with the sequential nature of the given data.


\subsection{Evaluation} \label{xval}
The built models are evaluated using cross validation, where data is repeatedly split into non-overlapping train and test sets. This way models are trained on one set, and afterwards tested on a test set comprised of unseen data to give a more robust estimate of the achieved generalization. However, special care needs to be taken when dealing with time series data.
Classical cross validation methods assume observations to be independent. This assumption does not hold for time series data, which inherently contains temporal dependencies among observations. We therefore split the data into training and test sets taking temporal order into account to avoid data-leakage. More concrete, we employ Walk Forward Validation (or Rolling Window Analysis) where a sliding window of $t$ previous trading days is used to train the models, and where $t_{t+1}$ is used for the out-of-sample test prediction. Table \ref{tab:evaluation} shows an example of this method where $t_{i}$ denotes the feature vector corresponding to trading day $i$. Note that in this scenario, when given a total of $n$ observations and a sliding window of length $t$, you can construct a maximum of $n-t$ different train-test splits.

\begin{table}[!ht]
 \caption{Example of Walk Forward Validation where $t_{i}$ represents the feature vector of trading day $i$. In this example, a sliding window of size three is taken. We therefore consistently use the feature vectors of the previous three trading days to train a model (underlined) and subsequently test said model on the fourth day (bold). } 
  \centering
  \begin{tabular}{cc}
    \toprule
    Iteration & Variable roles\\
    \midrule
     1 & $ \underline{t_{1} \hspace{1.5mm} t_{2} \hspace{1.5mm} t_{3}} \hspace{1.5mm} \mathbf{ t_{4} } \hspace{1.5mm} t_{5} \hspace{1.5mm} \dotsi \hspace{1.5mm} t_{n} $ \\
     2 & $ t_{1} \hspace{1.5mm} \underline{ t_{2} \hspace{1.5mm} t_{3} \hspace{1.5mm} t_{4}} \hspace{1.5mm} \mathbf{t_{5}}  \hspace{1.5mm} \dotsi \hspace{1.5mm} t_{n} $ \\
     \vdots & \vdots \\
     m & $ t_{1} \hspace{1.5mm} \dotsi \underline{\hspace{1.5mm} t_{n-3} \hspace{1.5mm} t_{n-2} \hspace{1.5mm} t_{n-1}} \hspace{1.5mm} \mathbf{t_{n}} $ \\

    \bottomrule
  \end{tabular}
  \label{tab:evaluation}
\end{table}

As with any cross validation method, models have to be retrained during each iteration of the evaluation process. Note that in this case, we also have to standardize and perform feature selection again each iteration as to not introduce look-ahead bias. 

\section{Experimental results and discussion}
In this section we present our experiment methodology and findings from our study. First we investigated whether we can use machine learning algorithms to predict if an asset’s market implied volatility will have moved up or down by the end of the next trading day. Second, we examined whether the introduction of alternative data sources has any positive impact on predictive accuracy. To this end, we conducted an ablation study where the effectiveness of each of the different sources is investigated. Five different scenarios were considered in total, shown in Table \ref{tab:ablation}. In each scenario, only original and generated features of the listed data sources were considered for prediction. 

\begin{table}[!ht]
 \caption{Different scenarios in which only features of listed sources are considered. This includes original and generated features. } 
  \centering
  \begin{tabular}{cl}
    \toprule
    Scenario & Description\\
    \midrule
    1 & Market data \\
    2 & Google News counts, Wikipedia traffic \\
    3 & Market data, Wikipedia traffic \\
    4 & Market data, Google News counts \\
    5 & Market data, Wikipedia traffic, Google News counts \\
    \bottomrule
  \end{tabular}
  \label{tab:ablation}
\end{table}

Evaluation is done on a temporal ordered dataset of 486 feature vectors each corresponding to a different trading day. For each of the different scenarios (Table \ref{tab:ablation}), we evaluate the models on 106 different train-test splits obtained using Walk Forward Validation (Section \ref{xval}). We employed a sliding window of 379 trading days (78\% of total available data), where the next day was used as the out-of-sample test case. We therefore made 106 consecutive out-of-sample movement predictions for AAPL's market implied volatility. Table \ref{tab:cdist} shows the class distribution for up and down movements during this period.

\begin{table}[!ht]
 \caption{Class distribution of down and up movements over the testing period of 106 days. } 
  \centering
  \begin{tabular}{cc}
    \toprule
    \multicolumn{2}{c}{Dependent Class Distribution}                   \\
    \cmidrule(r){1-2}
    Down Movement & Up Movement \\
    \midrule
    59\% & 41\% \\
    \bottomrule
  \end{tabular}
  \label{tab:cdist}
\end{table}

The results from our ablation study using Logistic Regression, a rbf-kernel SVM and AdaBoost are presented in Table \ref{tab:ablationres}. Note that Table \ref{tab:cdist} suggests we are dealing with a relatively imbalanced class distribution. We therefore considered balanced accuracy, which is defined as the average of recall obtained on each class, as evaluation metric instead of raw predictive accuracy. 

\begin{table}[!ht]
 \caption{Summary of obtained balanced accuracy for different models and data source scenarios. Note that the models used were standard configurations from Sklearn's Python library.} 
  \centering
  \begin{tabular}{c|ccc}
    \toprule
    Scenario & Logistic Regression & SVM & AdaBoost \\
    \midrule
    1 & \textbf{63.3\%} & \textbf{64.2\%} & 53.7\% \\
    2 & 52.3\% & 50.5\% & 55.0\% \\
    3 & 52.8\% & 55.6\% & \textbf{63.0\%} \\
    4 & 51.8\% & 53.3\% & 49.0\% \\
    5 & 61.3\% & 59.1\% & 57.3\%\\
    
    \bottomrule
  \end{tabular}
  \label{tab:ablationres}
\end{table}

Remarkably, using only features from market data (Table \ref{data_acquisition}) yielded the best results for both Logistic Regression and the SVM. This is in contrast with work in \cite{PstSPUEMaODS} and \cite{SmodAMPUDDS}, where alternative data was found to have a positive impact on stock price movement prediction. However, concluding that alternative data sources contain no useful information for implied volatility movement prediction is not entirely accurate. For example, the Adaboost classifier performed best using market and Wikipedia features. Logistic Regression performed poorly for this scenario, suggesting there is at least some non-linear relationship between Wikipedia features and movements in implied volatility that the AdaBoost classifier was able to find. 
Google News count appears to be the least effective data source in our experiments. Although combined with market and Wikipedia features, all models achieved their second best score. 

Despite the fact that we have shown AAPL's implied volatility movements can be predicted to a certain extent, the results are somewhat disappointing for the case of alternative data sources. It would be interesting to verify this behaviour on bigger datasets comprised of more trading days, as the vast amount of features might introduce too much noise for the models to effectively generalize on the relatively small training sets.

\section{Conclusion and Future Research}
We have shown that we can predict, to a certain extent, whether AAPL's market implied volatility at the end of next day will have moved up or down. The best results were obtained by only using market data, excluding Google News counts and Wikipedia page traffic. However, the results in Table \ref{tab:ablation} suggests there are predictive relationships to be found in these alternative data sources. How to effectively exploit these relationships is an open question. 

One of the biggest hurdles for this research was the availability of options data. We only had two years of end-of-day options data, which greatly constrained the time span for our research. Given the abundance of available features, it would be interesting to further investigate this topic on larger datasets. It also constrained us in our potential prediction targets. As our options data only consisted out of end-of-day statistics, we could only investigate movements from day end to day end. Lastly, there's a wealth of other alternative data sources left untapped (see Section \ref{interplay_section}) that might further increase predictive accuracy. However, exploiting these alternative data sources is mostly a labour intensive task. 

Aside from increasing efforts on feature engineering by generating smarter and more features, target construction can be improved. We only considered binary prediction where any small increment or decrement represented a move. The introduction of another label representing neutral moves is therefore definitely something worth investigating. Lastly, we didn't fully exploit the capabilities of the employed models. As we didn't perform parameter optimization for the employed machine learning algorithms, there might be room for improvement. In addition, all three models assign probabilities to the labels they predict. It would therefore be interesting to investigate whether higher probabilities are correlated with precision and movement magnitude.  



    

\bibliographystyle{unsrt}  

\bibliography{references}
\end{document}